\documentclass[english]{article}
\usepackage[utf8]{inputenc}
\usepackage[T1]{fontenc}
\usepackage{babel}
\usepackage{amsmath}
\usepackage{graphicx}
\usepackage{fancyhdr}
\usepackage{xcolor}
\usepackage{subcaption}
\usepackage{./pythonhighlight}
\usepackage{hyperref}
\begin{document}

\title{Spatial-Temporal Dataset of COVID-19 Outbreak in China}

\author{Wenyuan Liu\textsuperscript{1{*,+}}, Peter Tsung-Wen Yen\textsuperscript{1{*,+}}, Siew Ann Cheong\textsuperscript{1},}

\maketitle
\thispagestyle{fancy}

1. School of Physical and Mathematical Sciences, Nanyang Technological University, 21 Nanyang Link, Singapore 637371, Singapore

{*}corresponding author(s): Wenyuan Liu (wenyuan.liu@ntu.edu.sg), Peter Tsung-Wen Yen (peteryen2017@gmail.com)

{+} equal contribution
\begin{abstract}

We present Coronavirus disease 2019 (COVID-19) statistics in China dataset: daily statistics of the COVID-19 outbreak in China at the city/county level \cite{our_github}.
For each city/country, we include the six most important numbers for epidemic research: daily new infections, accumulated infections, daily new recoveries, accumulated recoveries, daily new deaths, and accumulated deaths.
We cross validate the dataset and the estimate error rate is about 0.04\%.
We then give several examples to show how to trace the spreading in particular cities or provinces, and also contrast the development of COVID-19 in all cities in China at the early, middle and late stages.
We hope this dataset can help researchers around the world better understand the spreading dynamics of COVID-19 at a regional level, to inform intervention and mitigation strategies for policymakers.
\end{abstract}

\section*{Background \& Summary}

Starting in East Asia at the beginning of 2020, COVID-19 is by now a global pandemic.
At the time of writing there are more than 392,000 confirmed cases in more than 190 territories with no sign of slowing down.
To meet this great healthcare challenge of our time, we need to combine efforts from the medical, pharmaceutical, epidemiological, transport, and even political realms \cite{Tianeabb6105, Chinazzieaba9757}.
Thus far scientific efforts made to understand how COVID-19 spreads, are mainly based on numbers from China at the province level.
As the situation improves in China, we can derive a more complete and detailed picture of COVID-19 spreading in the territory.
This picture can help other countries develop their own strategies to combat the  coronavirus.
At the time of writing, there are several datesets on the spread of COVID-19.
One of the most popular dataset is maintained by the Center for Systems Science and Engineering in John Hopkins University \cite{JHU_github} and others derived from this \cite{another_github}.
These datasets are mainly at the country or province level.
However, for advanced modeling and prediction, data at a smaller scale, i.e. the city level is necessary. 	
Unfortunately, there is no high-quality publicly-available datasets at the city level and covering a whole region to the best of our  knowledge, which scientists from all over the world can use.
In reality, Chinese authorities announce the ongoing situation daily after 20 Jan 2020, although (1) cities and provinces put their daily reports only in their own homepages and in different formats, and (2) most of these reports are in Chinese. 
It is therefore difficult for scientists who cannot read Chinese to do any research based on these reports.
For our own research, and also to make the data more widely accessible, we collected all daily reports  available from the official websites, extracted the data and organized them in several .csv files.
Researchers can then use their favorite tools to analyze the data, and it is our hope that this dataset can help people understand and fight COVID-19 better.

\section*{Methods}

At the end of 2019, the novel coronavirus was first discovered in Wuhan City, Hubei Province, China.
Since then, the viral infection spread out to nearby provinces and eventually to the whole of China.
Starting from 21 Jan 2020, provincial authorities have decided to release new and accumulated infected cases, newly recovered and accumulated recovered cases, death tolls, and other information to the public daily.
This information is published on the official Health Commission websites of each province once or twice a day (for some rare cases, we also found them reporting three to four times a day) depending on whether the infection situation is changing rapidly.
There are 22 provinces, 5 autonomous regions, 4 municipalities, two special administrative regions (Hong Kong and Macao), and also Taiwan.
In these official COVID-19 reports, cases are reported down to the administrative region level (the equivalent of a county).
For example, in Hubei Province, there are 17 administrative regions, such as Wuhan City, Huangshi City, Shiyan City, Yichang City, Xiangyang City, and so on.

Here, let us describe the procedures we used to extract essential information from the COVID-19 daily reports.
First, to ensure the reliability and verifiability of our data, we used a browser tool called \textit{Save Page WE} to download all the daily reports and save them locally as html files.
For consistency, the html files are named in the format “Province\_dd-mm-yyyy.html”.
We organized these source files into folders named after the provinces or regions.
\begin{figure}[ht]
	\centering
	\includegraphics[width=0.5\textwidth]{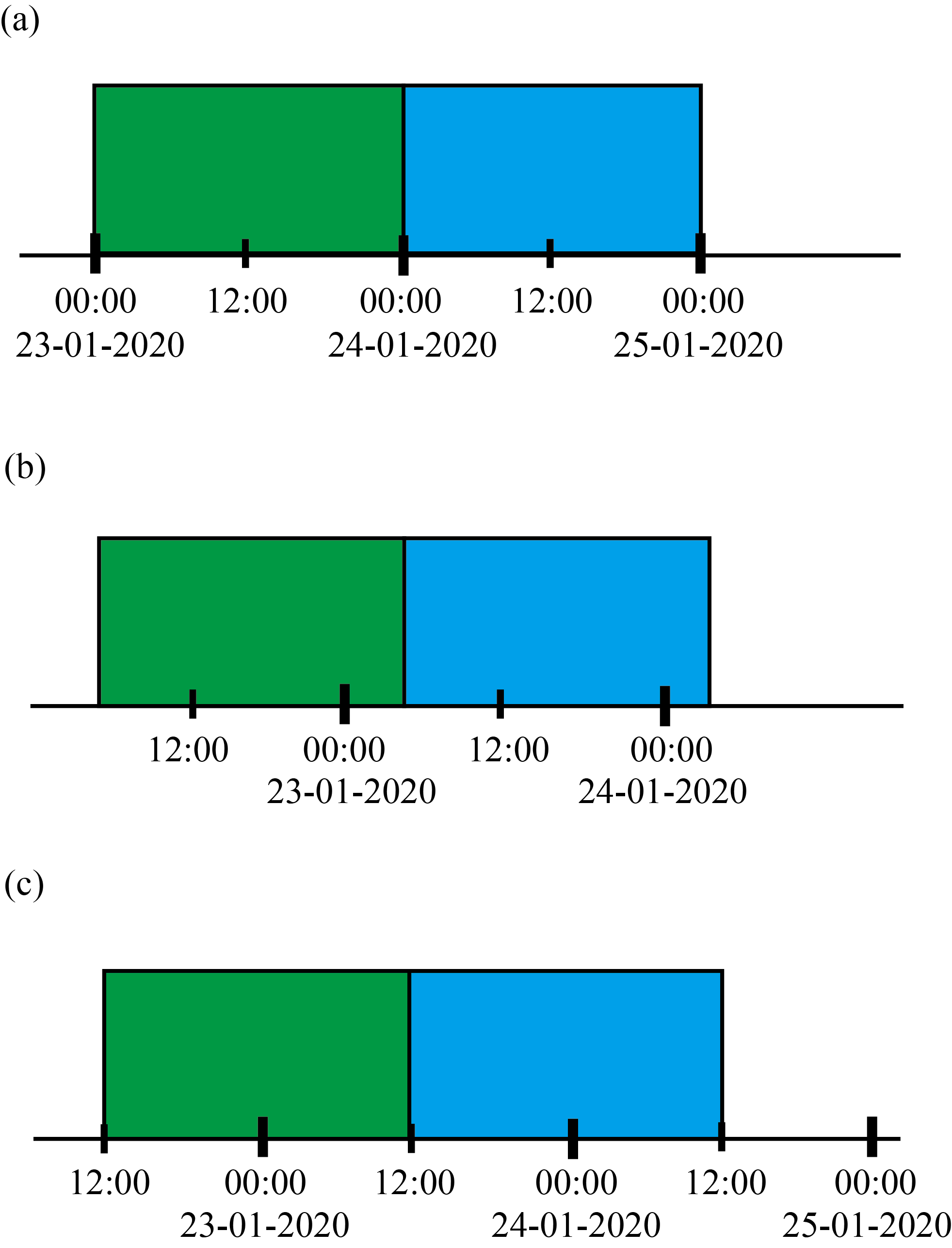}
	\caption{A schematic diagram illustrating cases having different time durations in the daily reports. The green and blue rectangles represent the coverage of two consecutive daily reports. (a) The duration is entirely within one calandar day. (b) The duration straddles two calendar days, and has uneven coverage. (c) The duration is split evenly between the two calander days.}
	\label{fig:report_timing}
\end{figure}
The whole dataset will be made accessible to all readers.
Next, we describe what information the daily reports reveal.
A typical daily report contains a duration of time, say for example 1600 hour on 23 Jan 2020 to 0900 hour on 24 Jan 2020.
If this duration is within one calandar day, we treat new cases reported therein as for that calendar day (see for example  Figure~\ref{fig:report_timing}(a)), whereas if the duration straddles two calendar days, we treat new cases reported therein as belonging to the calendar day with the longer coverage (see for example Figure~\ref{fig:report_timing}(b)); in cases where the duration is split evenly between the two calander days, we count new cases reported therein towards the earlier day (see for example Figure~\ref{fig:report_timing}(c)).
In the current version, we only extract (1) new and accumulated infected cases, (2) new and accumulated recovered cases, and (3) new and accumulated death cases, so we end up with six types of data for each of the administrative regions.
Because of their larger populations, municipalities report cases down to the district level.
Since municipalities are similar in sizes to counties, we decided to collect aggregated data for the municipalities so that our data set is uniform geographically.
Some provinces like Liaoning Province offer only aggregated data, and do not go down to the administrative region level.
For these cases, we collect and show only aggregated data.
Here we introduce three formulas that we used to count on day $i$ the new infected ($NI_i$), recovered ($NR_i$), and deaths ($ND_i$) from the accumulated infected ($TI_i$), recovered ($TR_i$), and deaths ($TD_i$):
\begin{equation}
NI_i = TI_i - TI_{i-1},
\end{equation}

\begin{equation}
NR_i = TR_i - TR_{i-1},
\end{equation}

\begin{equation}
ND_i = TD_i - TD_{i-1}.
\end{equation}
After we extract the reported cases from the daily reports, we use the above formulas to deduce the number of new cases for our dataset.

\begin{figure}[!hbt]
	\centering
	\includegraphics[width=\textwidth]{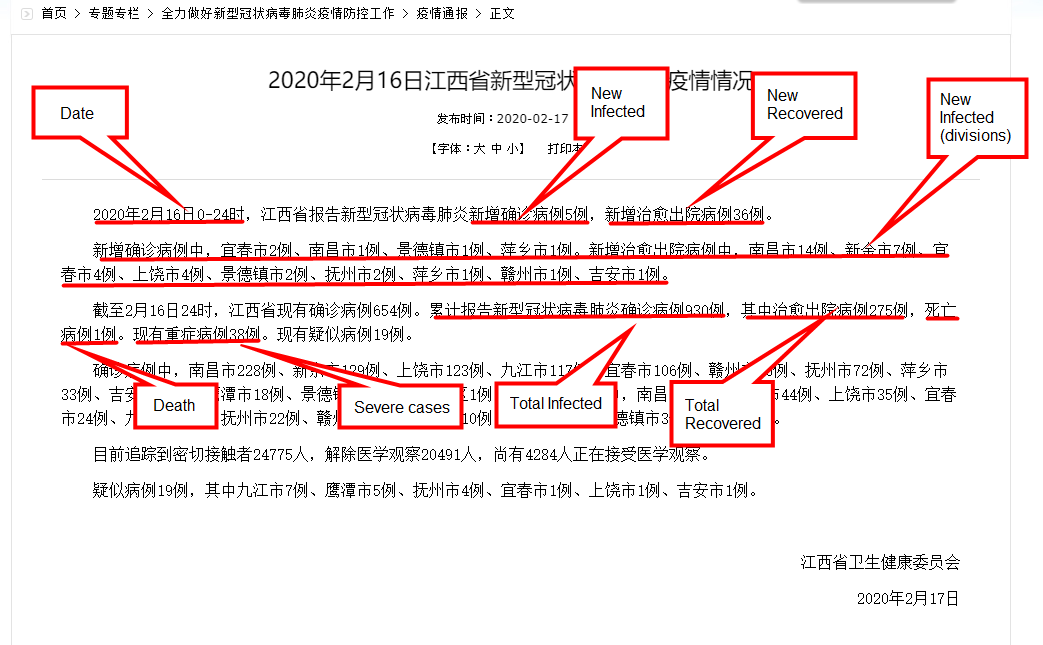}
	\caption{The COVID-19 raw data of Jiangxi Province. We used red boxes to indicate the data we extracted and stored in the dataset.}
	\label{fig:Jiangxi}
\end{figure}

\begin{figure}[!hbt]
	\centering
	\includegraphics[width=\textwidth]{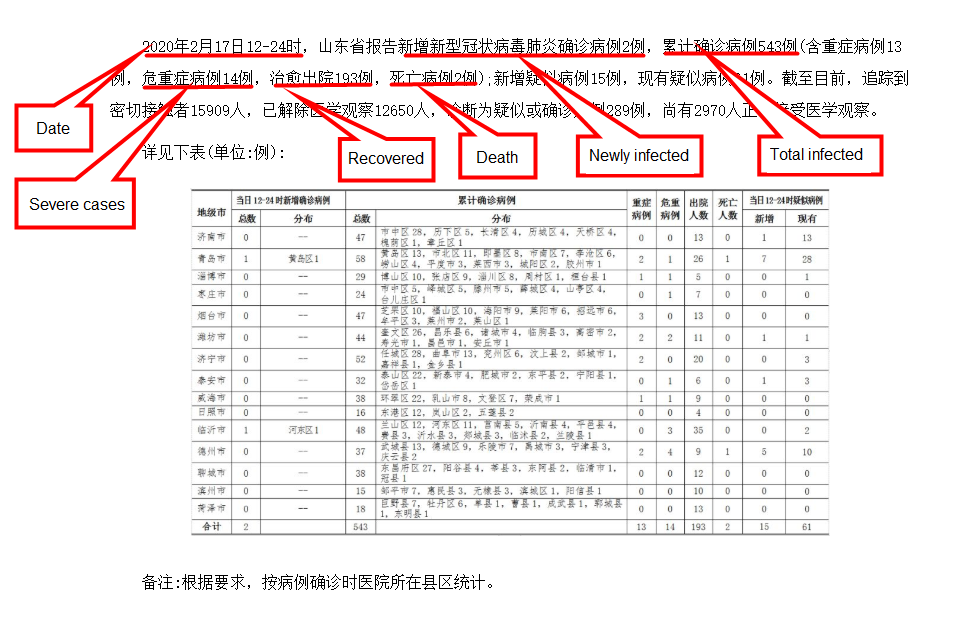}
	\caption{The COVID-19 raw data for Shangdong Province.}
	\label{fig:Shangdong}
\end{figure}

\begin{figure}[!hbt]
	\centering
	\includegraphics[width=\textwidth]{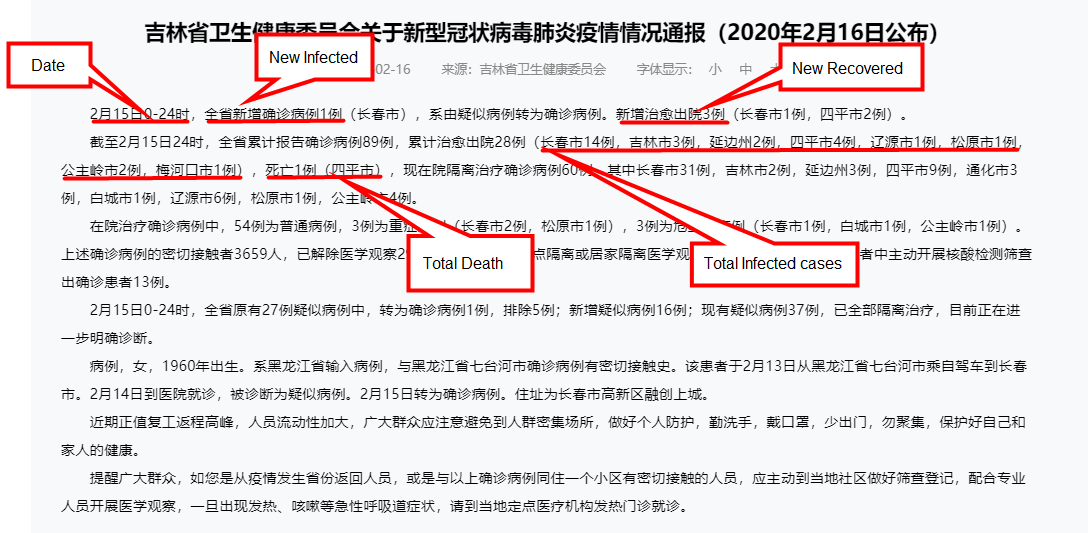}
	\caption{The COVID-19 raw data for Jilin Province.}
	\label{fig:Jilin}
\end{figure}

As a side note, in the early stage of this project, we planned to do data extraction automatically.
Unfortunately this was not successful because the report format for each province was different, making scripting approaches impractical and unreliable.
We show in Figures \ref{fig:Jiangxi}, \ref{fig:Shangdong}, and \ref{fig:Jilin} to illustrate how different the formats can be and the level of difficulty to automate the collection process. Nonetheless, we will continue to explore ways to make the automation procedure feasible in the future.


\section*{Data Records}
We made the dataset available through Github \cite{our_github} under the Creative Commons Zero v1.0 Universal (CC0-1.0) license.
We provide six .csv files to cover basic information of the COVID-19 pandemic in China,  namely the number of new confirmed infections (\textbf{China\_daily\_new\_infections.csv}) , the number of accumulated confirmed infections (\textbf{China\_accumulated\_infections.csv}), the number of new recovered patients (\textbf{China\_daily\_new\_recoveries.csv}), the number of accumulated recovered patients (\textbf{China\_accumulated\_recoveries.csv}), the number of new death case(s)  (\textbf{China\_daily\_new\_deaths.csv}), and the number of accumulated death case(s)  (\textbf{China\_accumulated\_deaths.csv}) on each day in each city.

Each file contains 368 lines and 44 columns: the first row is the header, the name for each column, while other rows are the data for all cities/counties.
For each row, the first four columns are names for city/county: the first cloumns is the name of city/county in English, the second column is the name of the provincial-level region this city/county belongs to in English, the third column is the name of city/county in Chinese, and the fourth column is the name of the provincial-level region this city/county belongs to in Chinese.
The remaining columns are dates ranging from 20 January 2020 to 29 Febraruy 2020 (in YYYY-MM-DD format).
For example, in \textbf{China\_accumulated\_infections.csv}, for row 169, column 1 is `Wuhan', whereas, column 19 (2020-02-04) is 8351.
This tells us that there are 8351 confirmed cases reported in Wuhan up till 24:00, 4 Febraruy 2020.

\section*{Technical Validation}

In the data collection process, we separated the 22 provinces, 5 autonomous regions, 4 municipalities, 2 special administrative regions (Hong Kong and Macao), and also Taiwan into two groups. 
The first two authors were then responsible for the extraction of data from daily reports and then convert them into spreadsheet files from each of the groups.
After completing this task, the two authors swapped the data groups and proceeded to do a cross validation on each other’s datasets. 
The purpose of swapping the datasets at this stage is to make sure that we eliminate as many of the  possible confirmation biases that can occur during the data extraction process. 
During this validation stage, we identified random errors like input errors, typos, and also registration errors (data consistently wrong after some dates). 
The registration errors can seriously degrade the quality of our datasets. 
For the 4 municipalities, the 2 special administrative regions, and Taiwan, the error rates are close to zero because they are not extracted at a city level.
However, in some of the provinces, like Hubei Province and Jiangsu Province, the error rates are as high as 5.7\% and 4.6\% respectively, which make these datasets unreliable without the cross validation. 
On average, we found that the error rate over the whole of our dataset is around 2\%. Therefore, after cross validation the error rate should be 0.04\%. 
Based on this estimate, we expect to find 34 errors in our 84,000-point dataset. This translates to finding one error for each province. Notwithstanding this, we believe our COVID-19 data set is robust and reliable enough after cross validation for other scientist to use for their respective rigorous studies. 
Next, we demonstrate step-by-step how to perform simple visualization and operations tasks on our COVID-19 dataset using Python.

\section*{Usage Notes}
Our datasets are in the common csv format, therefore researchers can use any software or programming language they prefer.
For example, to plot daily new infections and daily new recoveries in Wenzhou using the Python packages Pandas and Matplotlib, we can use the following code:

\begin{python}
import pandas as pd

import matplotlib.pyplot as plt

new_infections = pd.read_csv("China_daily_new_infections.csv")

total_infections = pd.read_csv("China_accumulated_infections.csv")

Wenzhou = total_infections[total_infections \\
['Prefectural level or Country level'] == "Wenzhou"]

dates = Wenzhou.columns[4:].tolist()

numbers = Wenzhou.iloc[0].tolist()[4:]

fig, ax = plt.subplots()

ax.bar(dates, numbers)

plt.xticks(dates, dates, rotation='vertical')

plt.title("Wenzhou")
\end{python}

\begin{figure}[!hbt]
	\centering
	\includegraphics[width=\textwidth]{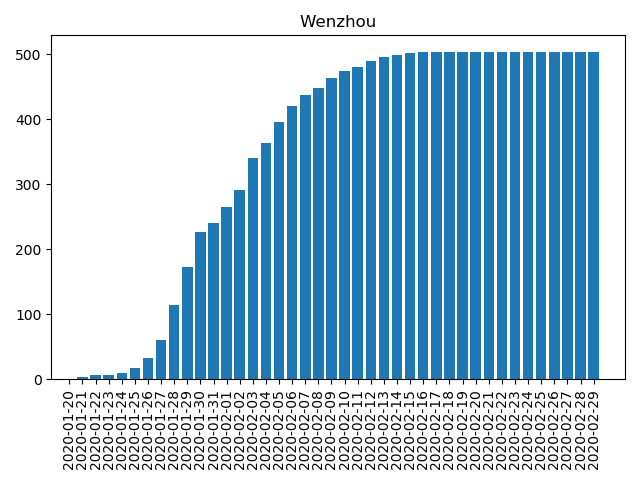}
	\caption{The accumulated number of COVID-19 infections in Wenzhou.}
	\label{fig:wenzhou}
\end{figure}

Beyond the city level, we can also study COVID-19 at the province level with our datasets by using the convenient "groupby" function in Pandas, as shown in the following Python code and figure.

\begin{python}

province_data = new_infections.groupby('Provincial-level regions').sum().reset_index()

dates = province_data.columns[1:].tolist()

Anhui = province_data[province_data['Provincial-level regions'] == "Anhui"]

Guangdong = province_data[province_data['Provincial-level regions'] == "Guangdong"]

fig, ax = plt.subplots()

ax.plot(dates, Anhui.iloc[0][1:].tolist(), label="Anhui")

ax.plot(dates, Guangdong.iloc[0][1:].tolist(), label="Guangdong")

ax.set_xticklabels(dates, rotation='vertical')

plt.legend()

\end{python}

\begin{figure}[!hbt]
	\centering
	\includegraphics[width=\textwidth]{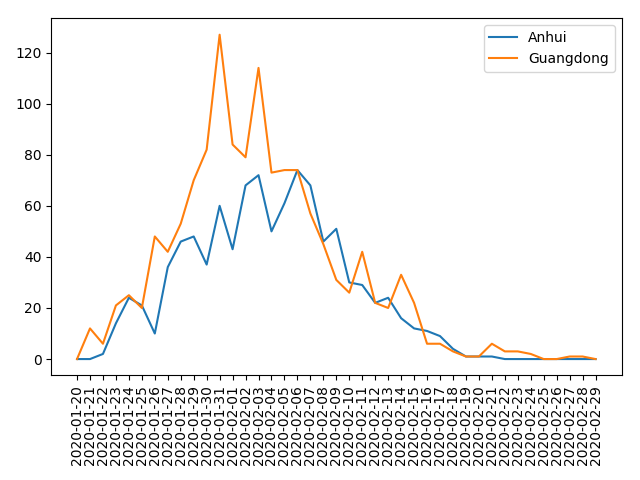}
	\caption{The comparison of new infection numbers of COVID-19 in Anhui and Guangdong.}
	\label{fig:anhui_and_guangdong}
\end{figure}

Through the use of a software like QGIS, our datasets can also be used to visualize the spreading of COVID-19 in whole of China with city-level resolution.

\begin{figure}[!hbt]
	\begin{subfigure}{\textwidth}
		\centering
		\includegraphics[width=1.2\linewidth]{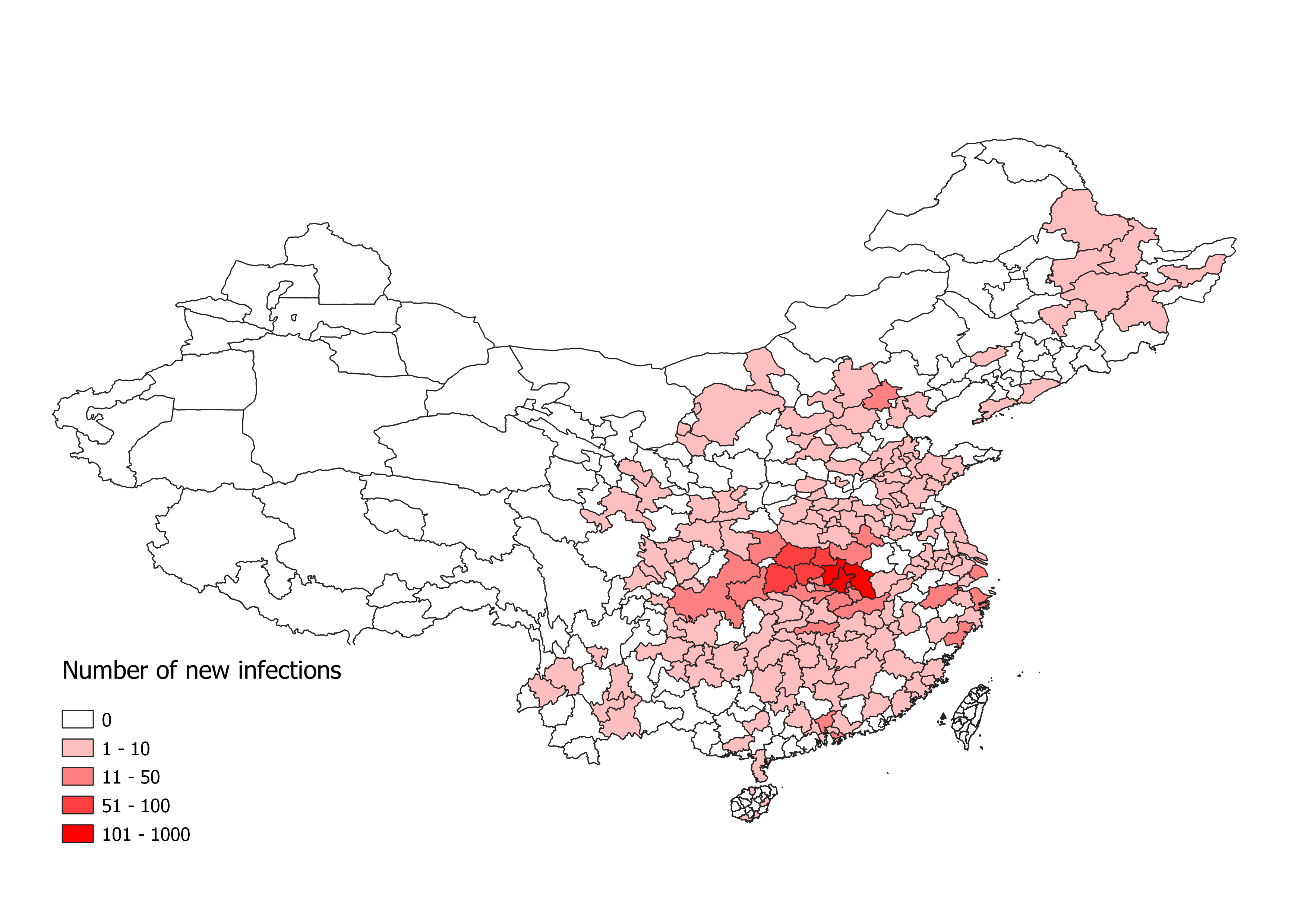}
		\caption{}
		\label{fig:sfig1}
	\end{subfigure}
	\begin{subfigure}{\textwidth}
		\centering
		\includegraphics[width=1.2\linewidth]{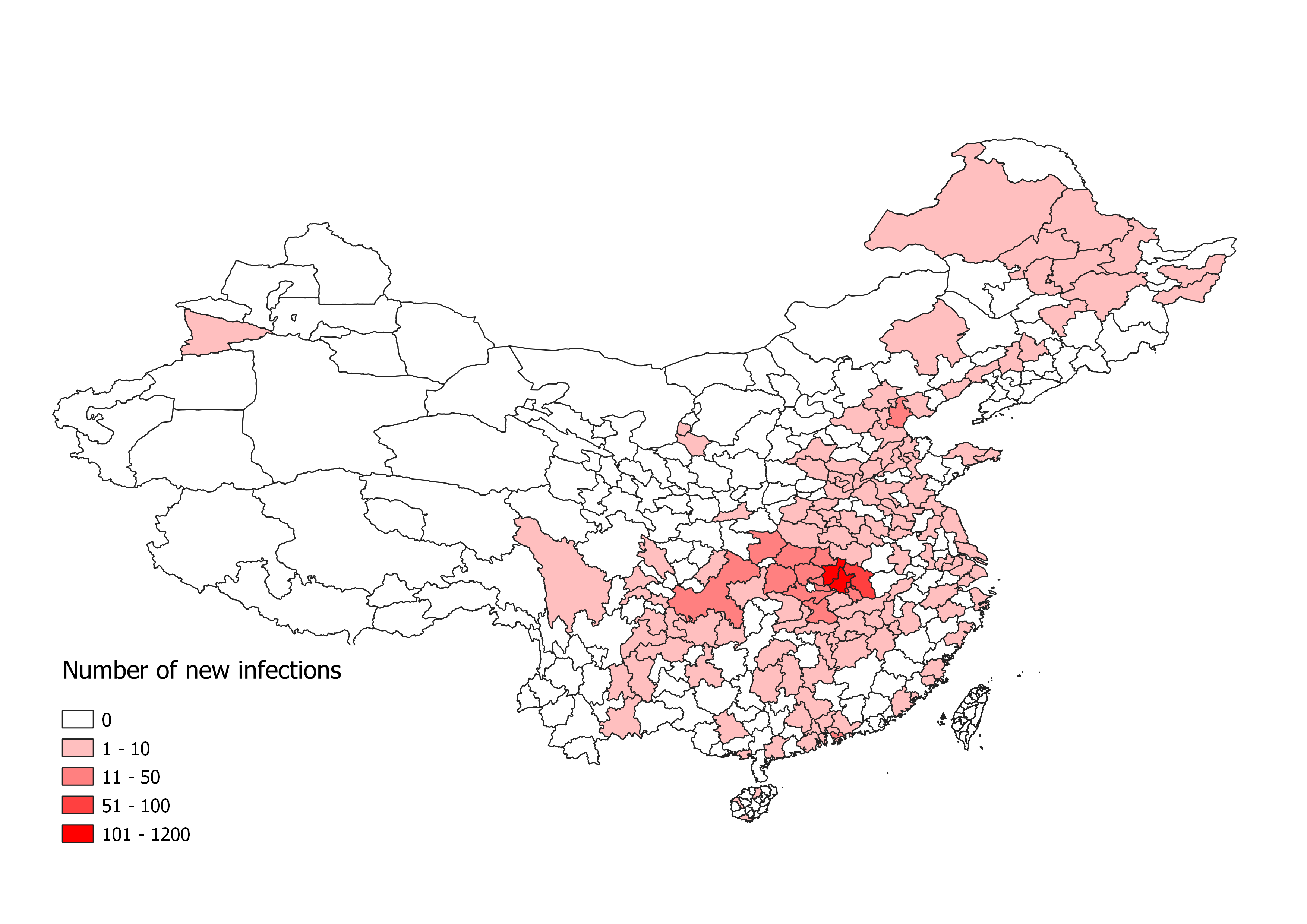}
		\caption{}
		\label{fig:sfig2}
	\end{subfigure}
\end{figure}

\begin{figure}
	\ContinuedFloat
	\begin{subfigure}{\textwidth}
	\centering
	\includegraphics[width=1.2\linewidth]{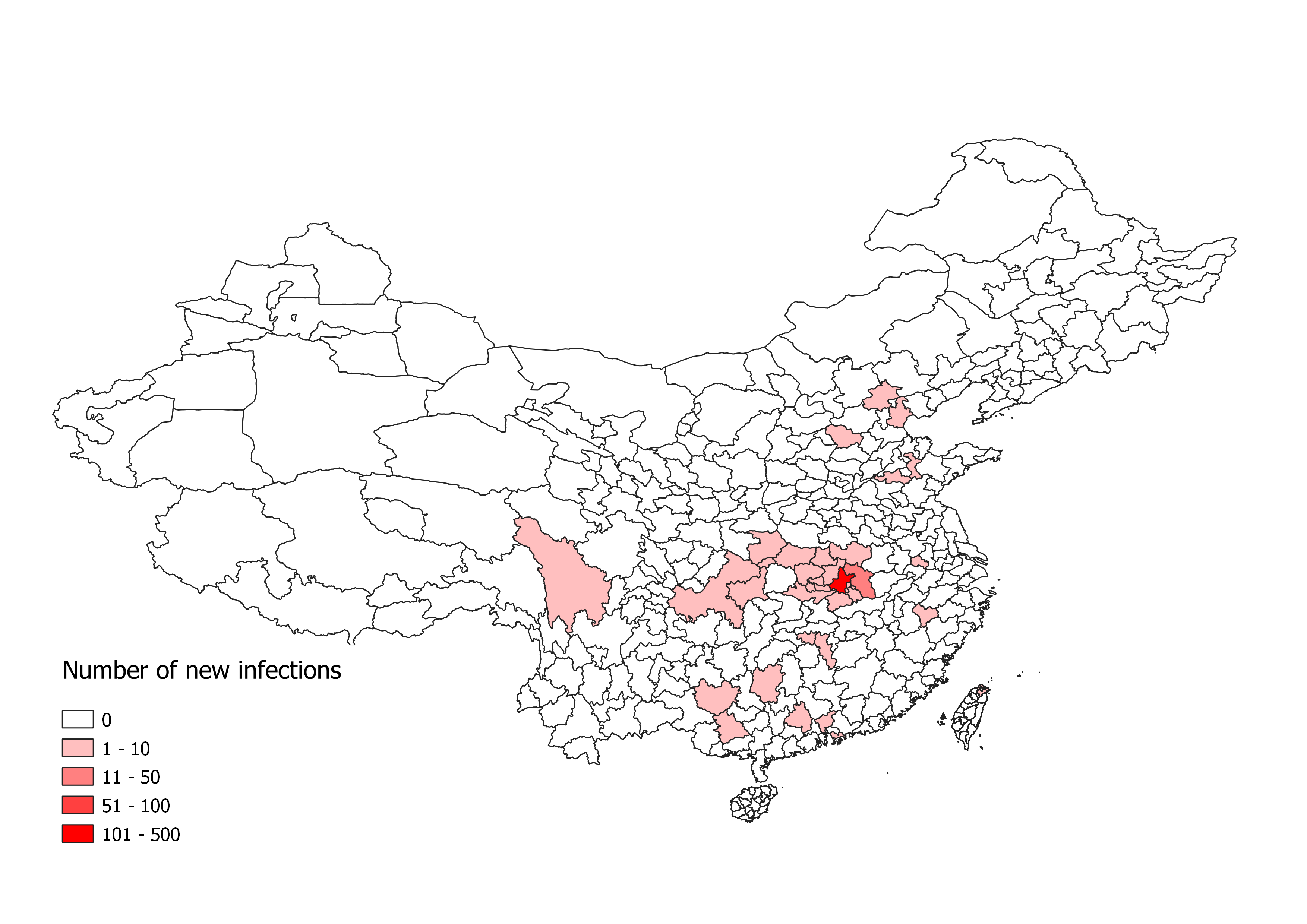}
	\caption{}
	\label{fig:sfig3}
	\end{subfigure}
	\caption{The number of new infections of COVID-19 of all cities in China on (a) 1 Feb 2020, (b) 11 Feb 2020, (c) 21 Feb 2020.}
	\label{fig:early_middle_late}
\end{figure}

\clearpage

\bibliographystyle{unsrt}
\bibliography{ref}

\end{document}